\documentclass[useAMS,usenatbib]{mn2e}

\usepackage[dvips]{graphicx}
\usepackage{amssymb}
\usepackage{amsmath}

\title[UV lines in gamma-ray binaries]{Tracing colliding winds in the UV line orbital variability of gamma-ray binaries}

\author[A. Szostek, G. Dubus and M. V. McSwain]{Anna Szostek$^{1,2}$\thanks{E-mail: aszostek@slac.stanford.edu}, Guillaume Dubus$^{3}$ and M. Virginia McSwain$^{4}$\\
$^{1}$Kavli Institute for Particle Astrophysics and Cosmology, Stanford
University, Stanford, CA 94305, USA\\
$^{2}$Astronomical Observatory, Jagiellonian University, Orla 171, 30-244 Krak\'ow, Poland\\
$^{3}$UJF-Grenoble 1 / CNRS-INSU, Institut de Plan\'etologie et d'Astrophysique de Grenoble (IPAG) UMR 5274, Grenoble, F-38041, France\\
$^{4}$Department of Physics, Lehigh University, 16 Memorial Drive E, Bethlehem, PA 18015, USA
}

\date{Submitted 2011}

\pagerange{\pageref{firstpage}--\pageref{lastpage}}\pubyear{2009}

\begin{document}

\maketitle

\label{firstpage}

\begin{abstract}
Gamma-ray binaries emit most of their radiated power beyond $\sim 10$ MeV.  The non-thermal emission is thought to arise from the interaction of the relativistic wind of a rotation-powered pulsar with the stellar wind of its massive (O or Be) companion star. A powerful pulsar creates an extended cavity, filled with relativistic electrons, in the radiatively-driven wind of the massive star. As a result, the observed P Cyg profiles of UV resonant lines from the stellar wind should be different from those of single massive stars. 

We propose to use UV emission lines to detect and constrain the colliding wind region in gamma-ray binaries. We compute the expected orbital variability of P Cyg profiles depending upon the interaction geometry (set by the ratio of momentum fluxes from the winds) and the line-of-sight to the system. We predict little or no variability for the case of LS 5039 and PSR B1259-63, in agreement with currently available HST observations of LS 5039. However, variability between superior and inferior conjunction is expected in the case of LS I+61 303.
\end{abstract}


\section{Introduction}

There are now five gamma-ray binaries detected in high (HE, $>$ 100 MeV) and very high energy (VHE, $>$ 100 GeV) gamma rays: PSR B1259-63, LS 5039, LS I+61 303, HESS J0632+057 and 1FGL 1018.6-5856. All contain a massive star and a compact object, and emit most of their power at energies above 10 MeV (for recent reviews see e.g. \citealt{paredes2011, hill2010}). The observed orbitally modulated HE and VHE gamma-ray emission indicate that particles are accelerated to multi-TeV energies within or close to the binary. The available observations in radio, X-rays and gamma-rays seem to favor a scenario in which particles are accelerated in the wind of a young pulsar \citep{maraschi1981,dubus2006} rather than in the jet of a microquasar \citep{romero2005,paredes2006,dermer2006}. The pulsar model is known to be operating in case of PSR B1259-63  \citep{tavani1994,Kirk1999}. The latest searches for radio pulsations in LS 5039 and LS I +61 303 yielded no positive result \citep{mcswain2011}, thus the presence of a wind collision region and pulsar remains to be proven in all remaining gamma-ray binaries.

\begin{figure*}
\center{\includegraphics[width=130mm]{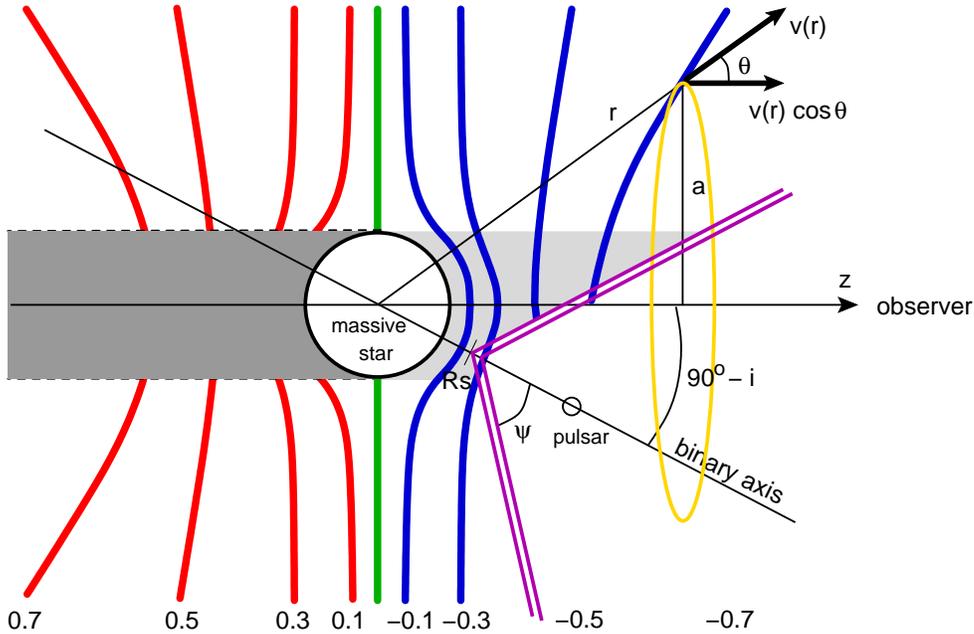}}
\caption{Section through a generic gamma-ray binary, cut along the binary axis and perpendicular to the orbital plane. Observer is to the right at infinity and sees the binary with an inclination $i \simeq 60^{\circ}$. Solid curves represent surfaces of constant wind velocity component towards the observer. The velocities in units of $v_{\rm z}/v_{\infty}$ are marked below each curve. Negative velocities are in blue, positive in red and green is zero velocity. Light gray shaded area marks the extent of the absorption zone. Dark grey area is the occulted wind region, which does not contribute to the final profile. Double solid line corresponds to the CD with $\eta < 1$.  The area to the right of the CD, is the pulsar wind cavity, filled with relativistic electrons which do not take part in resonant scattering of UV photons. The yellow ellipse is a cross section of one of the emission surfaces $S_{\nu}$.}
\label{f:scheme0}
\end{figure*}

A common denominator to all known gamma-ray binaries is a massive companion star (O or Be type). 
The massive star loses mass via a strong, hot, radiatively-driven wind at a rate that can be higher than $10^{-7}$ M$_{\odot}$ yr$^{-1}$ \citep{howarth1989}. UV resonant spectral lines formed in the wind are indicators of its velocity and mass-loss rate. If the wind density is large enough, the lines have a P Cygni profile consisting of a violet-shifted absorption component and red-shifted emission component. The P Cyg profiles form via absorption and scattering of photospheric continuum emission in the expanding, spherically symmetric stellar wind (see Sect. \ref{s:pcygni}).

The P Cyg line profiles from massive stars in a binary system with another massive star (e.g., O+O or WR+O) differ from those of a single star and are variable with orbital phase. The difference is caused by dynamical interactions between the two stellar winds and radiative feedback (e.g., wind ionization, radiative braking) between stellar winds and radiation field of the companion star. The variability of P Cyg profiles in massive star binaries was modeled numerically by \cite{stevens1993}. The results showed that P Cyg profiles have high diagnostic value for learning on the existence and the parameters of the colliding winds region.

UV line variability is also expected in high-mass X-ray binaries where a X-ray source ionizes the surrounding wind regions creating an extended Str\"omgren zone. The fully ionized plasma does not have any spectral transition, therefore the zone does not take part in resonant scattering, and the observed P Cyg line profiles should be different than in case of single stars. This effect was predicted by \cite{hatchett1977} and studied in detail by \cite{vanloon2001}.

We propose that line profile variability could be observed in the UV spectra of gamma-ray binaries within the framework of the pulsar wind scenario, where the relativistic wind of the pulsar creates a cavity in the companion stellar wind (see Sect. \ref{s:geometry}). The cavity is filled with relativistic particles which do not take part in resonant scattering of UV photons. Only the wind outside of the cavity is responsible for P Cyg profile. The cavity is described by its size (which increases with the parameter $\eta$ - see Sect. \ref{s:geometry} - and thus opening angle), and by location with respect to the line of sight. With an increasing size of the cavity, line profiles are expected to differ more from that of a single star. Concurrently, because the jets constitute narrow collimated outflows from the vicinity of the compact object, we do not expect the creation of the extensive cavity or significant wind alterations in the microquasar scenario. The presence or absence of line profile variability puts constraints on a size and location of the cavity. If changes in the line profile along the orbit are detected, a microquasar scenario can be rejected with high certainty.

In this article we study if and how the observed line profiles can vary in gamma-ray binaries in the framework of pulsar wind scenario. We constructed a semi analytical model composed of a massive star wind with a cavity. We describe the model in Sect. \ref{s:model}. In Sect. \ref{s:results} we show how the line profiles change with the shape and location of the cavity. In Sect. \ref{s:application} we apply the results to gamma-ray binaries LS 5039, PSR B1259-63 and LS I+61 303. In Sect. \ref{s:discussion} we discuss our results and present the conclusions.

\section{Model}
\label{s:model}
\subsection{Formation of P Cygni profiles}
\label{s:pcygni}
In most general terms, the P Cyg line profiles form via line scattering of photospheric photons in the hot expanding stellar wind of a massive star. The absorption line is formed when the photons emitted from a photosphere are scattered away from the line of sight by the stellar wind located in front of stellar disk and moving towards the observer (light grey area of Fig. \ref{f:scheme0}). The absorption component extends between Doppler velocities $-v_{\infty}$ and $0$. The emission line is created by the wind which surrounds the stellar disk (as seen by the observer) and which scatters the photospheric radiation into the line of sight, creating an emission line. If there was no wind, this radiation would never reach the observer. The emission line extends between Doppler velocities $-v_{\infty}$ and $v_{\infty}$. The radiation scattered behind the star (dark grey region of Fig. \ref{f:scheme0}), does not reach the observer. A superposition of the absorption and emission component creates the characteristic observed P Cyg profile \citep{lamers1999}.  

In more detail, the photons of frequency $\nu$ are scattered into or of the line of sight by wind elements moving with respect to the observer with a velocity $v_{\rm z} = v(r) \cos \theta = c(\nu-\nu_0)/\nu_0$ where $\nu_0$ is a rest frame frequency of a given resonant atomic transition. All the wind elements moving with the same velocity $v_{\rm z}$, are distributed along a surface $S_{\nu}$. The shape and emission of each $S_{\nu}$ are symmetrical with respect to the line of sight. A cross section across several of such constant velocity surfaces, corresponding to different frequencies, is shown in Fig. \ref{f:scheme0} as solid, blue, red and green lines. 

A cavity created by a pulsar wind may crop some of the constant velocity surfaces, breaking the axial symmetry and changing the flux received by the observer from the surfaces. The level of modification depends on the shape and location of the pulsar wind cavity with respect to the observer.

\subsection{The shape and location of a cavity}
\label{s:geometry}

The cavity is bounded by the colliding wind region, composed of two termination shocks separated by a tangential contact discontinuity (CD, double solid line in Fig. \ref{f:scheme0}). In the following we ignore the presence of the shocked winds and identify the shape of the pulsar wind cavity with the shape of CD. The shape of the cavity can be approximated by a conical surface symmetric with respect to the binary axis. Its vertex is located at the stagnation point, where the ram pressures of the two winds balance:
\begin{equation}
p_{\star}= {\dot M v_{_{R_{\rm s}}} \over 4 \pi R_{\rm s}^2}={\dot E \over 4 \pi c  (s-R_{\rm s})^2}=p_{\rm p},
\label{eq:Rs}
\end{equation}
where $R_{\rm s}$ is the distance of the stagnation point from the star center, $p_{\star}$ and $p_{\rm p}$ are (respectively) the stellar and pulsar wind ram pressures at the stagnation point, $s$ is the binary orbital separation at a given orbital phase, $\dot{M}$ is the stellar wind mass loss rate, $\dot{E}$ is the pulsar spin-down power and $v_{_{R_{\rm s}}}=v(R_{\rm s})$ is the stellar wind velocity at the radial distance $R_{\rm s}$ from the star's centre. For massive star it is given by a $\beta$-velocity law \citep{castor1975}
\begin{equation}
v(r) = v_{\infty} \left(1- {R_{\star} r_0 \over r} \right)^{\beta},
\label{eq:vwind}
\end{equation}
where $r_0 = 1-(v_0/v_{\infty})^{1/\beta}$, $v_0$ is the initial wind velocity, $v_{\infty}$ is the wind terminal velocity, $R_{\star}$ is the stellar radius and $\beta\approx 1$. 

The size of the cavity can be parametrized by the asymptotic opening angle of the cavity $\psi$, which in the conical approximation equals to half the vertex angle. We use the phenomenological relation derived from hydrodynamical simulations by \cite{bogovalov2008},
\begin{displaymath}
\psi = 28.6(4-\eta^{2/5})\eta^{1/3}\ \  {\rm for}\ \ \eta \leq 1,
\end{displaymath}
\begin{equation}
\psi = 171.5 - 28.6(4-\eta^{-2/5})\eta^{-1/3}\ \  {\rm for}\ \ \eta \geq 1,
\label{eq:psi}
\end{equation}
where $\psi$ is measured in degrees and where 
\begin{equation}
 \eta =  {\dot E \over \dot M v_{_{R_{\rm s}}} c },
 \label{eta0}
 \end{equation}
is the ratio of pulsar to stellar wind ram pressures at $R_{\rm s}$. When $\eta < 1$ then $\psi < 90^{\circ}$, the stellar wind dominates over the pulsar wind and the wind-free cavity is created around the pulsar as in Fig. \ref{f:scheme0}. When $\eta >1$ then $\psi > 90^{\circ}$, the pulsar wind dominates over the stellar wind and the conical CD encloses the star, efficiently blocking the wind from expanding into space around the binary. In this case the volume of the cavity can be larger than the volume of the expanding stellar wind. For low enough $\eta$, the stellar wind may also be prevented from freely expanding backwards by a reconfinement shock \citep{bogovalov2008}. Numerical simulations are then required to model the interaction region properly -- this is not taken into account here. Finally, no stable balance is possible when $\eta$ becomes very high: the pulsar wind then rams into the stellar surface \citep{Harding:1990gb}, quenching the stellar wind over at least part of one hemisphere.

For a given orbital phase, the location of the pulsar wind cavity can be parametrized using only $R_{\rm s}$ and the angle $\alpha$ between the line of sight and the binary axis 
\begin{equation}
\alpha(i,t,\omega) = \cos^{-1}\left[\sin i \cos\left(t-\omega\right) \right],
\end{equation}
where $t$ is the pulsar true anomaly (pulsar angular distance from periastron) and $\omega$ is the angle of the line of nodes of the orbit. Small values of $\alpha$ correspond to large inclination angles and orbital phases close to inferior conjunction (INFC, when the pulsar is in between the massive massive star and the observer). The maximum value is reached at superior conjunction (SUPC, when the pulsar is behind the star as seen by the observer) with $\alpha=180^{\circ}$ when $i=90^{\circ}$.

\subsection{Line profile calculations}
\label{s:lineprofile}
We followed \cite{lucy1971} (hereafter L71) to compute the line profile from the stellar wind of a massive star with a radius $R_{\star}$ and a stellar wind terminal velocity $v_{\infty}$. We model only single line profiles and do not consider line doublets where multi-line scattering takes place.  We assumed that the resonant transition $\nu_0$ is possible in the entire volume of the wind, as opposed to the situation where line transition is only possible in a certain wind layer due to the ionization gradient. We assume that the line opacity changes only due to changes in the gas density and do not take ionization effects into account. Here, we model line profile variability only for saturated profiles created in optically thick wind, but the qualitative results are also applicable to unsaturated lines from optically thin winds. 

To account for the presence of the pulsar wind cavity in the stellar wind, we modified the equation 43 of L71 which in its original form returns a normalized flux at frequency $\nu$ emitted by a corresponding constant velocity surface $S_{\nu}$. The modified equation 43 has the form:
\begin{displaymath}
F_{\nu_0}(\nu) ={2 \over R_{\star}^2} \int_{R_{\star}}^{\infty} \left( f_{\nu}(r) \Phi_{\nu}(\kappa,r,\nu) + \left[1-f_{\nu}(r)\right] \Phi_{\nu}(0,r,\nu)\right) \times 
\end{displaymath}
\begin{equation}
H(1-\mu^2) \times {r \over v(r)} {\partial v_{\rm z} \over \partial z} r dr
\label{e:mod43}
\end{equation}
where we assumed that the continuum flux equals 1, $\mu = \cos\theta$, $H$ is the Heaviside function, $\Phi$ is given by equation 44 in L71, whereas the last term by equation 42 in L71. 
$\kappa$ is an absorption coefficient set to an arbitrary value which gives saturated line profiles. The integral in Eq. \ref{e:mod43} includes a function $f_{\nu}$ which is a fraction of a circular cross section of $S_{\nu}$ (at $r$ perpendicular to the line of sight), that is outside of the pulsar wind cavity. An example cross section of $S_{\nu}$ for which $v_{\rm z} = -0.7v_{\infty}$ is shown as yellow ellipse in Fig. \ref{f:scheme0}. The function $f_{\nu}(r)$ depends on the shape of the cavity via $\eta$, and its location with respect to the observer via $R_{\rm s}$ and $\alpha$. $f_{\nu}(r)$ is calculated numerically for each binary configuration and frequency. 

In order to quantify the difference between P Cyg profile $P(v)$ of the same resonant line $\nu_0$ but for winds with different cavities, we introduce the profile area $A$, defined as:
\begin{equation}
A(\eta,R_{\rm s},\alpha) = \int_{-v_{\infty}}^{v_{\infty}} |P(v)-1| dv
\end{equation}
which is simply a sum of flux within the emission line above continuum and flux removed from continuum by absorption. In the following we will use $A_0$ to denote a profile area normalized by a profile area of a single star. 

A spectral resolution of COS instrument onboard of Hubble Space Telescope in the 900 -- 1450 \AA\ wavelength range and using bright object aperture, is $R = 3200-4200$ \citep{cos2011}. This corresponds to 40--90 km s$^{-1}$ (2--6\% of $v_{\infty}$ depending on spectral type). Assuming that the line is fully saturated, we estimate that for this resolution, a change in profile can be detected if a ratio of profiles' areas changes at least by 0.05-0.1.

\begin{figure}
\center{\includegraphics[width=75mm]{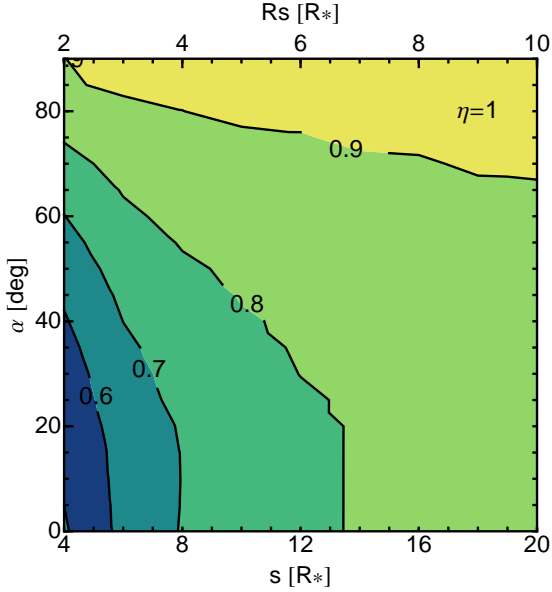}}
\caption{Contour plot of the normalized profile area $A_0$ as a function of the binary separation $s$ and the angle $\alpha$ (between the cavity axis and line of sight). Here, $\eta=1$ is fixed, which implies $\psi=90^{\circ}$ and $R_{\rm s}=0.5s$ are also fixed. Line profile increasingly resembles that of a single star ($A_0$=1) with increasing $s$ and $\alpha$.}
\label{f:separationvsalpha}
\end{figure}
\begin{figure}
\begin{tabular}{c}
\includegraphics[width=77mm]{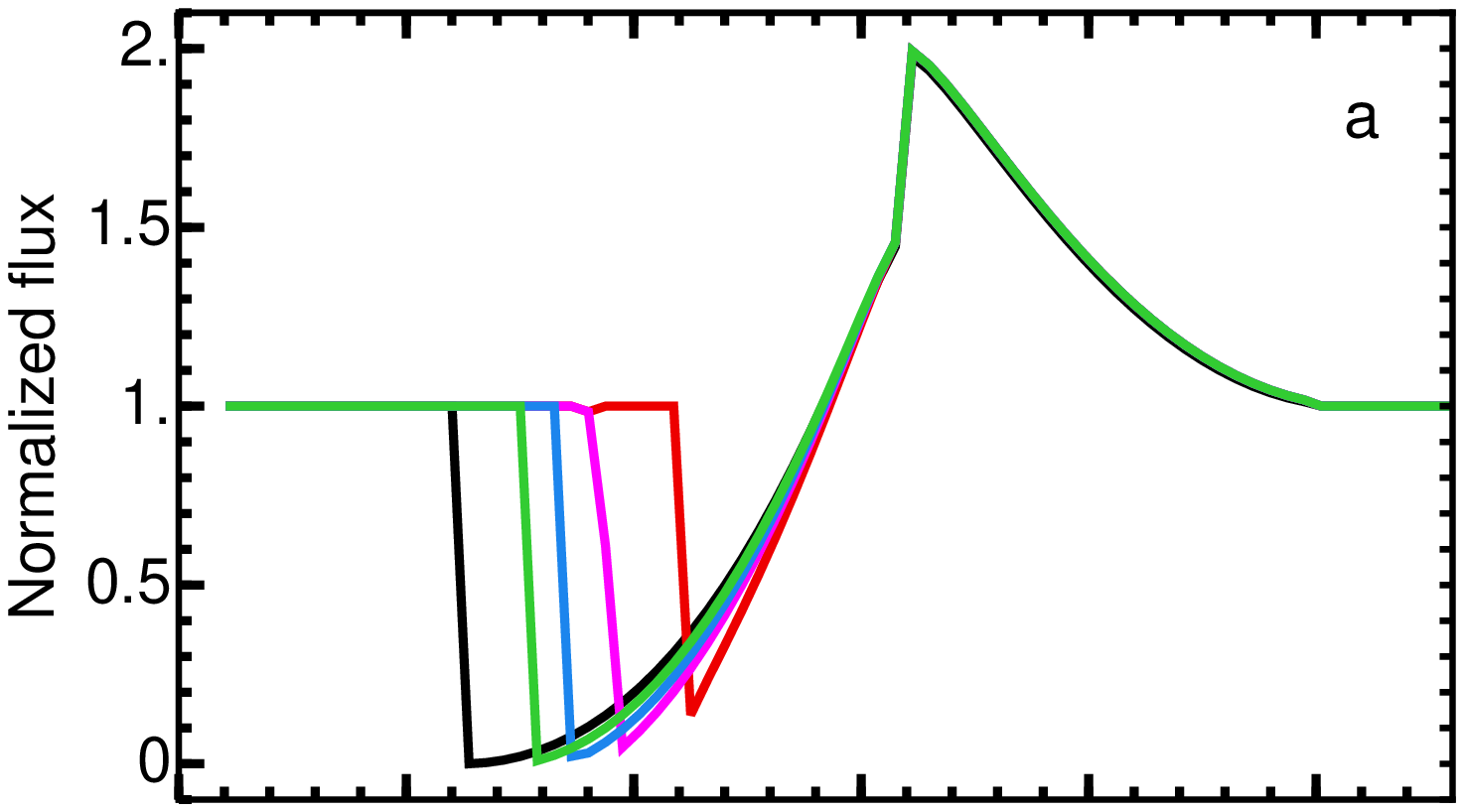}\\[-2.1mm]
\includegraphics[width=77mm]{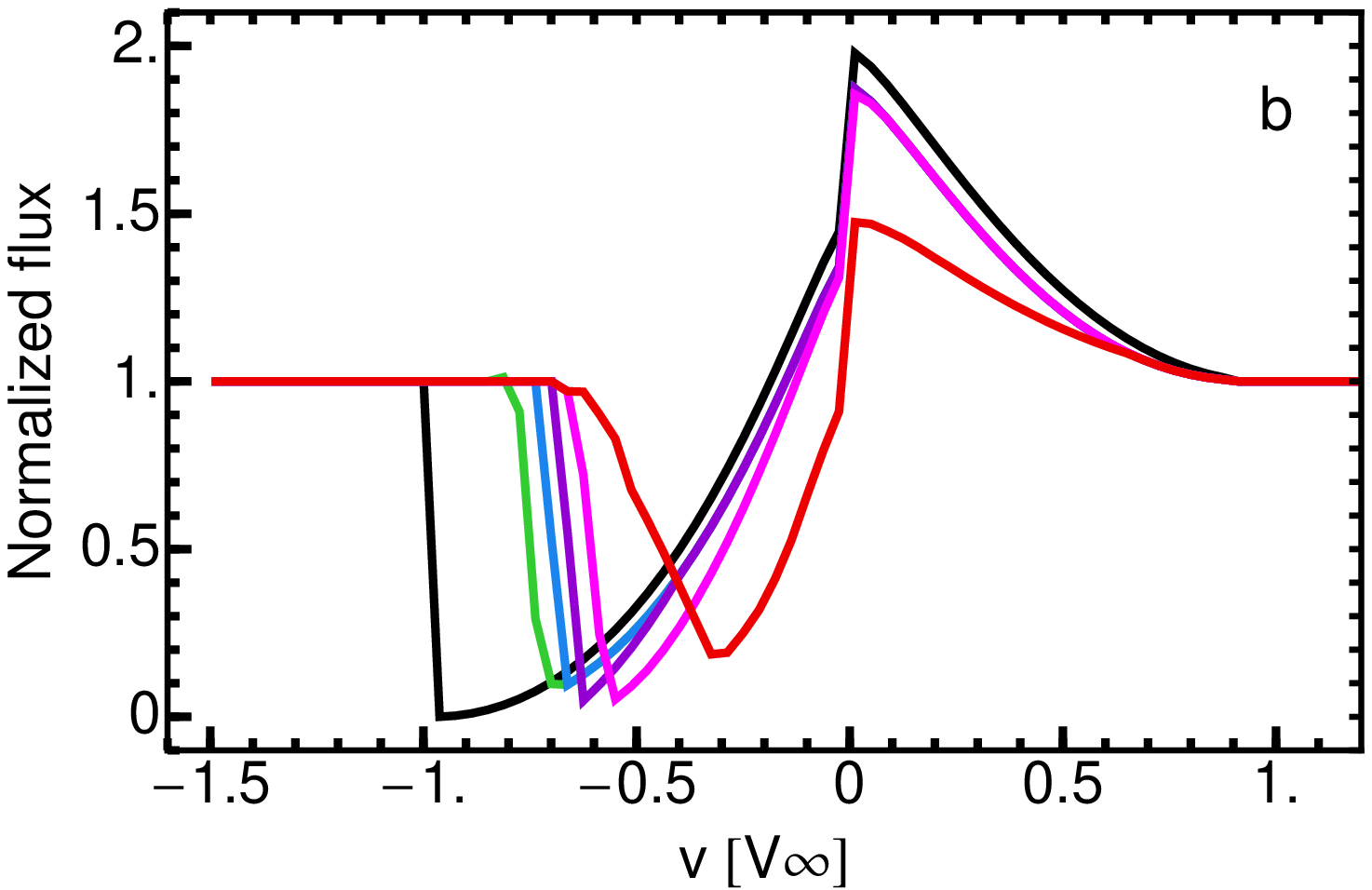}\\
\end{tabular}
\caption{P Cyg profiles for different binary configurations. Top: $\eta=1$, $\alpha=0$ are fixed and the binary separation $s$ is changed. Profiles shown $s=\infty$ (single star, black, $A_0=1$),  $s = 4R_{\star}$ (red, $A_0=0.6$), $s = 6R_{\star}$ (magenta, 0.7), $s = 8R_{\star}$ (blue, 0.8) and $s = 13R_{\star}$ (green, 0.9). Bottom: $R_{\rm s} = 3R_{\star}$, $\alpha=0$ are fixed and $\eta$ is changed (equivalent to varying cavity opening angle $\psi$). Profiles correspond to $\eta = 0.02 $, $\psi = 30^{\circ}$ (green, $A_0=0.72$); $\eta = 0.23$, $60^{\circ}$ (blue, 0.68); $\eta = 1$, $90^{\circ}$ (violet, 0.65); $\eta = 4.5$, $120^{\circ}$ (pink, 0.65); $\eta = 50$, $150^{\circ}$ (red, 0.5).}
\label{f:lightcurves}
\end{figure}
\begin{figure}
\center{\includegraphics[width=75mm]{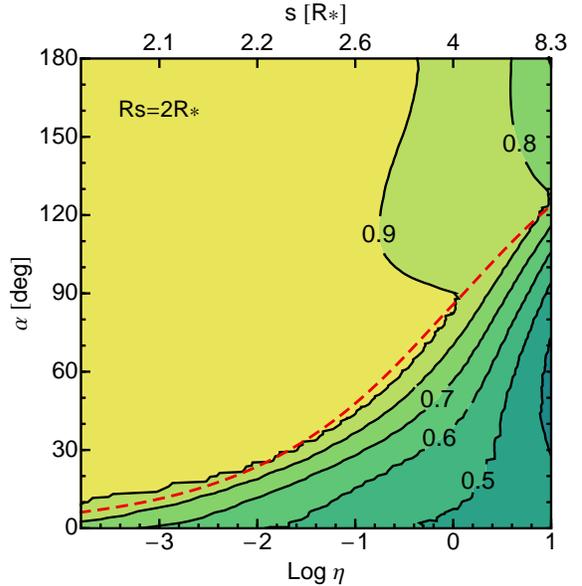}}
\caption{Contour plot of the normalized profile area $A_0$ as a function of $\eta$ and $\alpha$ with  $R_{\rm s} = 2R_{\star}$ fixed. The corresponding binary separation $s$ is plotted at the top of the figure. The red dashed line shows $\alpha=\psi$.}
\label{f:etavsalpha}
\end{figure}
\begin{figure}
\center{\includegraphics[width=75mm]{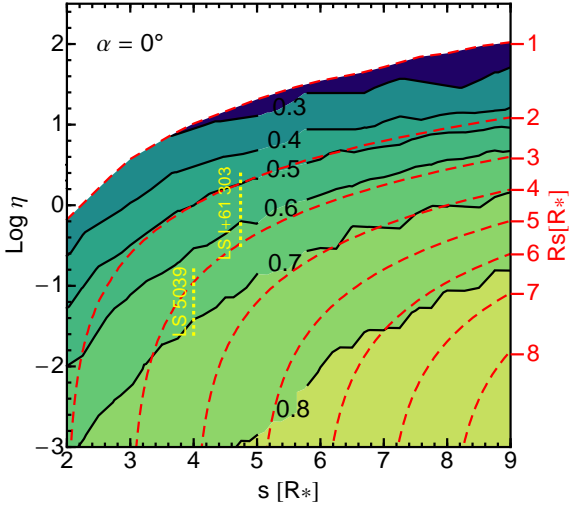}}
\caption{Contour plot of the normalized profile area $A_0$ as a function of binary separation $s$ and $\eta$ for $\alpha=0^\circ$ fixed. Constant values of $R_{\rm s}$ in this parameter space are shown as dashed lines. The white region at the top of the plot is where the wind interaction region collides with the stellar surface. The expected parameter range for LS 5039 and LS I+61 303 is also shown.}
\label{f:etavsseparation}
\end{figure}
%


\section{Results}
\label{s:results}

\subsection{Parameter survey}
We chose to study the variations in line profiles by changing $\eta$, $\alpha$ and $s$ (or $R_{\rm s}$). A set of three parameters is  sufficient to describe all possible configurations. We successively freeze one of the parameters and show contour plots of $A_0$ as a function of the other two. Note that the physical length scale is set by $R_\star$.

Figure \ref{f:separationvsalpha} shows $A_0$ as a function of $s$ and $\alpha$ when $\eta=1$, which implies $\psi=90^{\circ}$ and $R_{\rm s}=0.5s$. The line profile from a gamma-ray binary differs more from that of a single star when the cavity is closer to the star (smaller $R_{\rm s}$ in units of $R_\star$ or, equivalently, smaller $s$, implying shorter  orbital periods) and when the axis of the cavity is close to the line of sight (low $\alpha$). Example profiles for $\alpha=0$ are shown in Fig. \ref{f:lightcurves}a. When the cavity moves in closer to the star, the stellar wind is prevented from reaching high velocities in the absorption zone (Fig. 1). Photons that would have been  scattered away from the line of sight can then reach the observer unaltered, which explains the narrower absorption trough. For $\eta=1$, the red emission component of the line profile barely changes since it arises from regions unaffected by the cavity.

Figure \ref{f:etavsalpha} shows $A_0$ as a function of $\eta$ and $\alpha$ when $R_{\rm s}= 2R_{\star}$. Note that keeping $R_{\rm s}$ fixed implies that  the pulsar power increases with $\eta$ (hence $s$), up to $\dot E \sim 10^{38}$ erg s$^{-1}$ when $\eta=10$ and assuming $\dot M = 10^{-7}$ M$_{\odot}$ yr$^{-1}$. The line profile is increasingly modified when $\eta$ becomes high ($=$ wider cavity opening angle $\psi$). The line profile is most sensitive to the presence of the cavity when $\alpha < \psi$ i.e. when the line-of-sight passes through the cavity for some orbital phases. For $\alpha>\psi$, the profile is only weakly sensitive to changes of $\alpha$ or $\eta$. When $\alpha<\psi$, changes in the blue absorption line (created by the wind located along the line-of-sight) are responsible for the strong variations in $A_0$. Example profiles are shown in Fig. \ref{f:lightcurves}b with $R_{\rm s} = 3R_{\star}$ and $\alpha=0^{\circ}$. The red emission part of the line is modified when $\psi$ (hence $\eta$) is very high, disrupting the wind well beyond the absorption zone.

Finally, Fig. \ref{f:etavsseparation} shows $A_0$ as a function of $\eta$ and $s$ when $\alpha$ is held fixed at zero. The general trends are in agreement with those shown in Fig. \ref{f:separationvsalpha}-\ref{f:etavsalpha}. $A_0$ decreases with increasing $\eta$ and decreasing $s$. The dependence on the separation $s$ is relatively weak. Instead, the $A_0$ contours follow closely the contours of constant $R_{\rm s}$, as expected based on the Fig. \ref{f:separationvsalpha}. Fig. \ref{f:etavsseparation} shows that deviations from a single star profile can be observed when $R_{\rm s} \lesssim 6R_{\star}$, assuming the most favorable binary configuration ($\alpha=0^\circ$). The white area at the top of the plot covers the parameter region where the wind collision region intercepts the stellar surface (see \S2.2).

These results are in very good agreement with the calculations of \cite{stevens1993} for binaries composed of two massive stars, in particular with their Model 1b. The profile variability shown in Fig. 2b of \cite{stevens1993}, can be directly compared with our Fig. \ref{f:lightcurves}a. Our calculations apply to any inclination, orbital phase and shape of the cavity.

\subsection{Survey limitations}
The line profile strongly depends on the cavity opening angle $\psi$. \cite{gayley2009} derived a relation between $\eta$ and the asymptotic value of $\psi$, that is different from that used here, based on other assumptions about the flow in the shocked region. The opening angles under these assumptions are larger from the ones given by Eq. \ref{eq:psi}. The difference is about $10^{\circ}$ for $0.1<\eta<1$ and increases to $\sim15^{\circ}$ for smaller $\eta=0.005$. Based on Fig. \ref{f:etavsalpha}, we estimate that in the most favorable case, when $\alpha=0^{\circ}$, the $15^{\circ}$ increase of $\psi$ leads to a decrease in $A_0$ of $\approx 0.1$, hence of minors consequences to our results. Improved accuracy in predicting line profiles is more likely to require advanced stellar wind models and numerical simulations of the pulsar wind interaction with stellar wind.

In our calculations we assumed a spherical symmetry of the pulsar wind momentum distribution. The simulations of pulsar winds from high velocity neutron stars interacting with the ISM provide insight on the possible impact of a latitudinal variation of pulsar wind on the shape of the interaction surface. The results of the simulations performed by \cite{vigelius2007} show, that while some changes in the structure of the interaction region close to the pulsar are possible, the overall geometry remains mostly the same and is determined principally by $\dot E/c$. The present study is far from the level of detail where it would be necessary to take these effects into account.

Three out of five known gamma-ray binaries, contain a massive Be companion star. The class of stars is known for their highly anisotropic winds composed of a fast, tenuous polar wind and a dense Keplerian equatorial disk. The interaction of the pulsar wind with the equatorial disk can cause strong variations in momentum ratio $\eta$ and disturbance of the global interaction surface \citep{okazaki2011}. The shape of the pulsar wind cavity during disk passage and its impact on the UV lines are uncertain and can not be addressed by our model.

\subsection{Application to known systems}
\label{s:application}
We apply our model to the gamma-ray binaries with the best known parameters: PSR B1259-63, LS 5039 and LS I+61 303. We compare line profiles at conjunctions (i.e. at minimum and maximum value of $\alpha$ for any given orbit) in order to test whether a pulsar wind cavity could be detected.

\subsubsection{PSR B1259-63}
This is the only gamma-ray binary known to contain a pulsar (spin-down power $\dot E = 8 \times 10^{35}$ erg s$^{-1}$). The companion is a  Be star, which has a fast, tenuous polar wind and a slow, dense equatorial disk wind. It is the polar wind that is responsible for the P Cyg profile. 

The polar wind mass-loss rate in PSR B1259-63 is about $\dot M_{\rm p} = 1 \times 10^{-7}$ M$_{\odot}$ yr$^{-1}$, following \cite{vink2000} using the stellar parameters from \cite{neguerela2011}. The corresponding value of $\eta$ is $\sim 10^{-2}$. For this value of $\eta$ and for $i \lesssim 25^{\circ}$ \citep{neguerela2011}, $\alpha > \psi$ and we do not expect to see changes in line profile (see Fig.~\ref{f:etavsalpha} and \S3.1). 

\subsubsection{LS 5039}
UV observations of LS 5039 at two orbital phases, 0.41 (close to INFC) and 0.63 (close to apastron), did not reveal any changes in the P Cyg line profiles \citep{mcswain2004}. We set constraints on the binary parameters from the lack of variability.

The value of $\eta$ depends on the balance between the stellar and (putative) pulsar wind, but the parameters of the latter are unknown in LS 5039. \cite{szostek2011} estimated a maximum value of $\eta_{\rm max}=0.6$ at periastron that, based on Eq. \ref{eq:Rs} and \ref{eq:psi}, corresponds to $\psi \simeq 75^{\circ}$ and $R_{\rm s} = 1.2 R_{\star}$. For higher values of $\eta$, the pulsar wind impinges directly on the stellar surface, disrupting the stellar wind over much the pulsar-facing side (the pulsar is only 1-3 stellar radii away from the surface) -- which is excluded by the observations. With $\dot M=10^{-7}$ $M_{\odot}$ yr$^{-1}$, $v_{\infty} = 2.4 \times 10^8$ cm s$^{-1}$, $R_{\star}= 9.3R_{\odot}$ \citep{casares2005} and $v_0 = 2 \times 10^6$ cm s$^{-1}$, $\eta_{\rm max}$ is reached for a pulsar with $\dot E \simeq 5 \times 10^{36}$ erg s$^{-1}$, which is higher than in PSR B1259-63 but not unknown among other pulsars \citep[e.g.,][]{kargaltsev2010}. \cite{zabalza2011} set a comparable upper limit from the absence of strong thermal X-ray emission from the shocked stellar wind. LS 5039 has a mildly eccentric orbit, meaning that $\eta$ and $\psi$ decrease and $R_{\rm s}$ increases at phases away from periastron. Hence, line profiles calculated with $\eta=\eta_{\rm max}$ at  periastron give an upper limit on the orbital  variability that can be expected. In this case, the cavity parameters at conjunctions are (SUPC) $\eta=0.38$, $\psi = 70^{\circ}$, $R_{\rm s} = 1.4R_{\star}$ and (INFC) $\eta=0.17$, $\psi = 55^{\circ}$ and $R_{\rm s} = 2.8R_{\star}$. The inclination of the binary is unknown: $i<60^{\circ}$ if the X-ray source is point-like \citep{casares2005} but $i$ can be as high as $90^{\circ}$ if the X-ray source is extended \citep{szostek2011}. However, the observed gamma-ray luminosity requires a pulsar spin-down power at least similar to PSR B1259-63, in which case $\eta\geq \eta_{\rm min} \approx 0.04$. 

\begin{figure}
\begin{center}
\begin{tabular}{c}
\includegraphics[width=77mm]{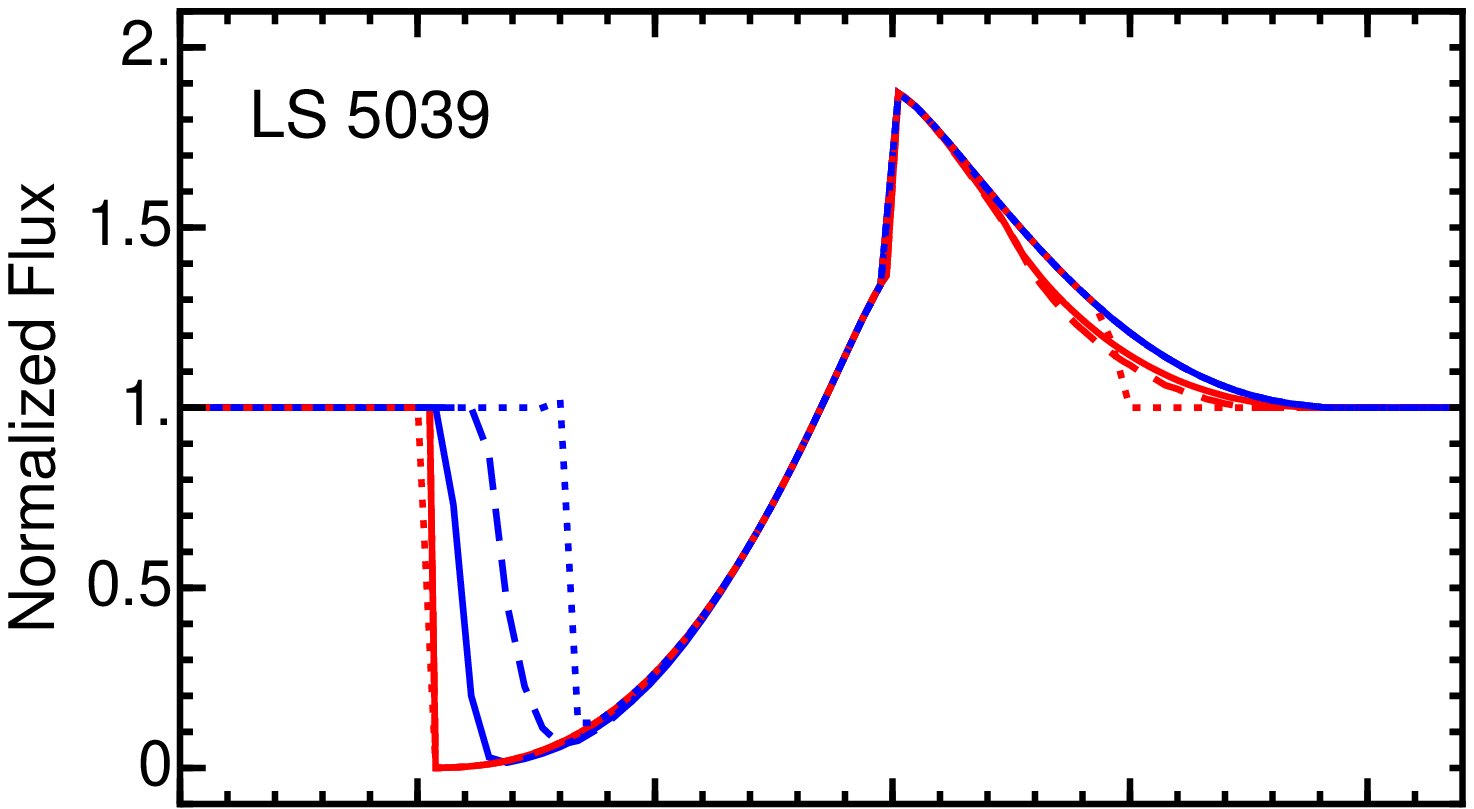} \\[-1.8mm]
\includegraphics[width=77mm]{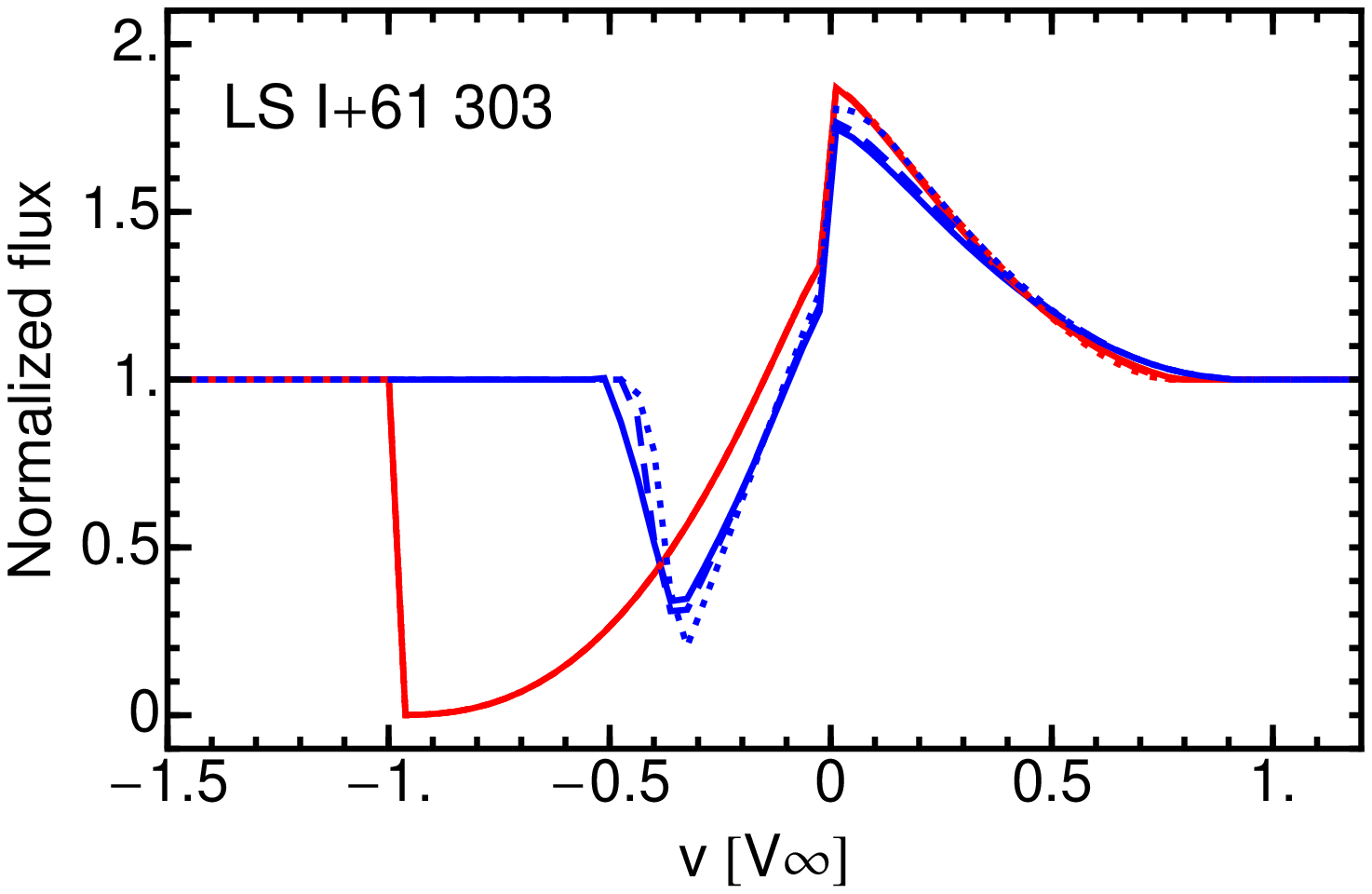} \\
\end{tabular}
\end{center}
\caption{Expected P Cyg profiles for LS 5039 and LS I+61 303 at two orbital phases (SUPC, red and INFC, blue) assuming $\eta=\eta_{\rm max}$ (see text for details) and inclination angles of $50^{\circ}$ (solid), $60^{\circ}$ (dashed), $90^{\circ}$ (dotted).} 
\label{f:binaryprofiles}
\end{figure}

The range of $\eta$ discussed above is marked in Fig. \ref{f:etavsseparation}, showing the expected change in line profile is modest  even for the most favorable case $\alpha=0^\circ$. The difference in P Cyg profiles (Fig. \ref{f:binaryprofiles} between SUPC and INFC for $\eta=\eta_{\rm max}$ should be detectable down to $i\approx 50^{\circ}$. From this result and the fact that line profiles of LS 5039 are constant as observed by HST, we conclude that $\eta$ must be much less than $\eta_{\rm max}$ or that $i<50^{\circ}$. There is a degeneracy between the two parameters but the higher the value of $i$, the lower $\eta$ and $\dot E$ need to be to avoid line variability. At high inclination, even small cavities with $\eta \sim 10^{-3}$ cause an observable ($A_0\leq 0.9)$ modulation of lines profiles along the orbit. If $\eta=\eta_{\rm min}$ then variability in the profiles is detectable  unless $i\leq70^\circ$. Therefore, we conclude that the absence of variability in the line profiles observed with HST requires low values of $\eta$ if the inclination is high, with an upper limit $i\leq 70^\circ$, with higher values of $\eta$ possible if the inclination is smaller.

%
%

%
%


\subsubsection{LS I+61 303}
LS I+61 303 also harbors a massive Be star. 
 \cite{romero2007} suggested that the pulsar wind would dominate over the polar wind ($\eta > 1$) over much of the orbit, creating an extended cavity in the stellar wind. Because the polar wind is less dense (lower $\dot M$) than the wind in LS 5039, the line profiles are expected to be weaker but observable. The IUE spectra of LS I+61 303 (generally low dispersion and poor S/N) obtained between 1978--1983 indicate that several emission lines formed in the wind show variability of uncertain nature \citep{howarth1983}. 

Taking the stellar parameters ($R_{\star} = 10 M_{\odot}$, $v_{\infty}= 10^8$ cm s$^{-1}$, $\dot M = 10^{-8} M_{\odot}$ yr$^{-1}$)  and the orbital solution derived by \cite{aragona2009}, we estimate $\eta_{\rm max}\approx 4.5$ at periastron, for which $R_{\rm s} = 1.3 R_{\star}$ and $\psi = 120^{\circ}$. The spin-down power implied is $\dot E \simeq 4 \times 10^{36}$ erg s$^{-1}$. The cavity parameters  change from (SUPC) $\eta = 1.5$, $R_{\rm s} = 4.4R_{\star}$ and $\psi=100^{\circ}$ to (INFC) $\eta=2.5$, $R_{\rm s} = 1.8 R_{\star}$ and $\psi=110^{\circ}$. The expected LS I+61 303 profiles at conjunctions are plotted in Fig. \ref{f:binaryprofiles}. In Fig. \ref{f:etavsseparation} we also indicate a range of $\eta$ for values of $\dot E$ between $8 \times 10^{35}$ and $4 \times 10^{36}$ erg s$^{-1}$ at $\alpha=0$. We expect to see a significant variability of line profiles along the orbit in LS I+61 303 because $\eta$ is always relatively high in this system for plausible values of $\dot{E}$. If the pulsar interacts with the Be disc instead of the polar wind at some orbital phases then $\eta$ is expected to be much lower than $\eta_{\rm max}$ due to the increased density, resulting in an increase of $A_0$. This may result in some of the unexplained UV variability although this is difficult to assess with the currently available data.


\section{Conclusions}
\label{s:discussion}

We modeled the line profile orbital variability due to the presence of a pulsar wind-generated cavity in a gamma-ray binary. The line profiles increasingly differ from those of a single star when the cavity is largest and located closer to the line of sight, that is for smaller $R_{\rm s}$, larger $\eta$ and smaller $\alpha$. The strong dependence on $\alpha$ means that significant variations are expected whenever the cavity intersects the wind region responsible for the blueshifted absorption component. The emission component of the P Cyg profile remains steady except for very large values of $\eta$ (and $\psi$).

We applied our model to the binaries LS 5039, PSR B1259-63 and LS I+61 303. We predict there is no observable line profile variability in PSR B1259-63 due to large separation along the orbit and small inclination angle. HST UV spectroscopy of LS 5039 did not show a change in line profiles between phases close to INFC and apastron \citep{mcswain2004}. This implies that the inclination is lower than 50$^\circ$ for the estimated maximum $\eta_{\rm max}=0.6$ and no larger than 70$^\circ$ if $\eta=0.04$. For lower values of $\eta$ the spin-down power of the pulsar is insufficient to explain the $\gamma$-ray luminosity. 

Our model calculations predict a significant line profile modulation in LS I+61 303 if $\eta>1$ as argued by \cite{romero2007}. The modulation should be observable even if the binary inclination angle is low. The shape of the cavity may be drastically changed during disk wind passage, which should be visible as the line profile returning to the single star profile. Monitoring of line profile variability with orbital phase  in LS I+61 303 could be used not only to learn about the existence and parameters of the pulsar wind cavity, but also to search for signatures of interactions with the equatorial wind of the Be companion. In addition, if changes in the line profile along the orbit are detected, a microquasar scenario can be rejected with high certainty.

\section*{Acknowledgments}
MVM is supported by NASA DPR number NNX11AO41G, National Science Foundation grant AST-1109247, and an institutional grant from Lehigh University. This work was supported by the European Community via contract ERC-StG-200911. 

\bibliographystyle{mn2e} 
\bibliography{P_Cyg} 

\end{document}